\documentclass[aps,pra,amsmath,amssymb,nofootinbib,floatfix]{revtex4}

\usepackage[T1]{fontenc}
\usepackage[T1]{tipa}
\usepackage{graphics}
\usepackage{physics}
\usepackage{bm}
\usepackage{times}
\usepackage{color}
\usepackage{natbib}
\usepackage{hyperref}
\usepackage{xcolor}
\usepackage{jabbrv}

\DeclareMathOperator{\sinc}{sinc}

\begin{document}

\title{Information encoding in the spatial correlations of entangled twin beams}

\author{Gaurav Nirala$^{1,2,}${\footnote{gauravnirala@ou.edu}}, Siva T. Pradyumna$^{1,2}$, Ashok Kumar$^{1,3}$ and  Alberto M. Marino$^{1,2,4,5,}$\footnote{marino@ou.edu, marinoa@ornl.gov}}
\affiliation{$^{1}$Homer L. Dodge Department of Physics and Astronomy, The University of Oklahoma, Norman, Oklahoma 73019, USA \\
$^{2}$Center for Quantum Research and Technology (CQRT), The University of Oklahoma, Norman, Oklahoma 73019, USA\\
$^{3}$Department of Physics, Indian Institute of Space Science and Technology,
Thiruvananthapuram, Kerala, 695547, India \\
$^{4}$Quantum Information Sciences Section, Oak Ridge National Laboratory, Tennessee, 37381, USA\footnote{This manuscript has been authored in part by UT-Battelle, LLC, under contract DE-AC05-00OR22725 with the US Department of Energy (DOE). The publisher acknowledges the US government license to provide public access under the DOE Public Access Plan (http://energy.gov/downloads/doe-public-access-plan).}\\
$^{5}$Quantum Science Center, Oak Ridge National Laboratory, Tennessee, 37381, USA}

\begin{abstract}
The ability to use the temporal and spatial degrees of freedom of quantum states of light to encode and transmit information is crucial for the implementation of a robust and efficient quantum network. In particular, the large dimensionality of the spatial degree of freedom promises to provide significant enhancements; however, such promise has largely been unfulfilled as the necessary level of control over the spatial degree of freedom to encode information remains elusive. Here, we show that  information can be encoded in the distribution of the spatial correlations of highly multi-spatial mode entangled bright twin beams. We take advantage of the dependence of the spatial correlations on the angular spectrum of the pump required for four-wave mixing, as dictated by phase matching. The encoded information can be extracted by mapping the momenta distribution of the twin beams to a position distribution in the far field and measuring the spatial cross-correlation of images acquired with a high quantum efficiency electron multiplying charge coupled device. We further show that the encoded information cannot be accessed through individual beam measurements and that the temporal quantum correlations are not modified. We anticipate that the ability to engineer the distribution of the spatial correlations will serve as a novel degree of freedom to encode information and hence provide a pathway for high capacity quantum information channels and networks. In addition, a high degree of control over the spatial properties of quantum states of light will enable real-world quantum-enhanced spatially resolved sensing and imaging applications.
\end{abstract}

\maketitle

The use of quantum correlations to enable new and enhanced real-world technology is one of the defining features of the ongoing second quantum revolution~\cite{dowling2003RSPTA}. Quantum correlations in the spatial degree of freedom, in particular, promise to have a significant impact on quantum information science due to the infinite dimensionality of the corresponding Hilbert space~\cite{kolobov1999RMP,Fabre20,boyd2015NJP,gisin2017PRL}. Specifically, quantum-correlated spatial modes of light can provide higher capacity information encoding down to the single photon level~\cite{zeilinger2016PNAS,zeilinger2001nature}, while simultaneously offering better security and robustness to noise in optical communication protocols~\cite{forbes2020OSA}. While for classical systems spatial modes of light have been already used for high-capacity transmission both in free-space and fiber based communication channels \cite{ashrafi2015AdOP,willner2012NatPhotonics,ramachandran2013Science}, in the quantum regime this has been an elusive task due to the limited control and manipulation of spatial correlations. Here, we go beyond the demonstration of spatial correlations and show that the ability to engineer their distribution can enable a novel degree of freedom to encode and transfer information efficiently. Moreover, this capability can extend beyond the efficient transfer of information on a quantum network into applications such as quantum-enhanced imaging and sensing, for which the presence of multiple spatial modes can lead to  enhanced resolution and sensitivity~\cite{genovese2020APL,gatti2008PRA,fabre2000OptL}.

Entangled twin beams of light are at the heart of a number of applications in quantum information science and are natural candidates for encoding information in the spatial degree of freedom as they can be generated with a large number of spatial modes~\cite{Boyer2008,Boyer2018Science,Holtfrerich16a}. Their generation requires a nonlinear parametric process, such as parametric down conversion (PDC) or four-wave mixing (FWM), that can simultaneously emit pairs of photons. Such a nonlinear parametric process conserves both the energy and momentum of the involved fields, with energy conservation leading to temporal quantum correlations and momentum conservation (or phase matching) to spatial quantum correlations.  As a result of phase matching, in particular, the spatial quantum properties of the generated photons depend on the spatial properties of the input pump photons~\cite{kolobov1999RMP,Fabre20}. This dependence provides an effective way to introduce quantum correlations between specific spatial modes of the output photons. The engineering of the correlations between spatial modes will give rise to structured quantum states of light~\cite{forbes2021NatPhotonics} with a tailored mode distribution for specific applications.

The relation between the spatial properties of the pump and the spatial quantum correlations of the generated photons has been extensively studied theoretically for PDC~\cite{walborn2010PhysRev}, with a few experiments showing that a pump with a modified spatial structure leads to a change of the spatial correlations between the generated photons~\cite{monken1998PRA,arnaut2002PRA,gigan2021OptL}. Similar results have been experimentally observed for FWM~\cite{Marino08,Holtfrerich16,jing2017OptLett,jing2017PRA,glasser2018OptLett}. While the degree of control over the spatial correlations has been limited in these previous experiments, they show the viability of using the pump to modify the distribution of the spatial correlations. Here, we show that it is possible to take advantage of this approach and use a structured pump beam with a specific angular spectrum (momentum distribution) to engineer the distribution of the spatial correlations of twin beams.  This makes it possible to encode information jointly in the twin beams, such that if a particular $k$-vector of one of the beams is measured the distribution of $k$-vectors in the other beam contains the encoded information.

\begin{figure}
	\centering
	\includegraphics{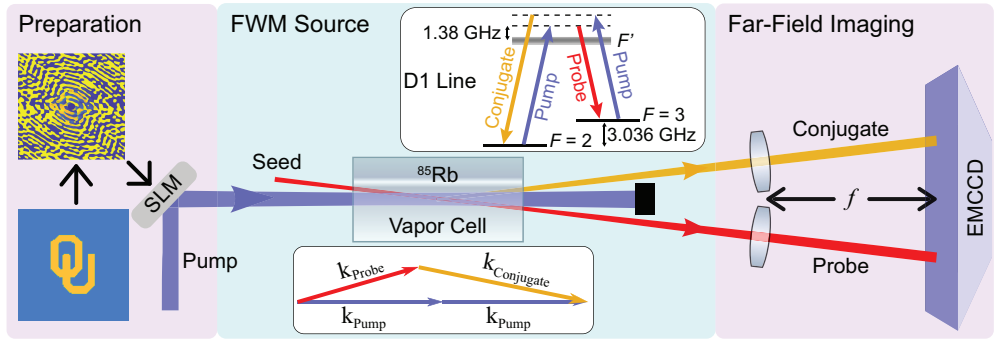}
	\caption{Experimental setup for encoding information in the distribution of the spatial correlations of twin beams. A hot $^{85}$Rb vapor cell is used as the nonlinear medium needed for the FWM process that generates quantum-correlated twin beams, which we refer to as probe and conjugate. The FWM is based on a double-$\Lambda$ configuration in the D1 line of $^{85}$Rb, as shown on the top inset. The pump beam is reflected from an SLM that imprints a phase pattern onto it to obtain the necessary momentum distribution (angular spectrum) for the pump. The phase-structured pump is then imaged to the center of the cell via a 4$f$ optical system.  Finally, the momentum distribution of the probe and the conjugate beams is mapped to a position distribution onto an EMCCD camera in the far field using a $f$-to-$f$ imaging system. Images acquired with the EMCCD are then used to measure the distribution of the spatial correlations and extract the encoded information in the twin beams. In order to generate bright twin beams, we seed the FWM with an input probe beam to achieve a photon flux of $\sim 10^{14}$ photons/s per output beam, which is limited by saturation of the EMCCD.  SLM - spatial light modulator; EMCCD - electron multiplying charge coupled device.}
	\label{fig:setup}
\end{figure}

In our experiments, we use a FWM process based on a double-$\Lambda$ configuration~\cite{McCormick2008} (top inset in Fig.~\ref{fig:setup}) to generate twin beams, which we refer to as probe and conjugate. This process has been shown to generate quantum states of light that are entangled in the temporal~\cite{Boyer2018Science} and spatial~\cite{ashok2021IOP} domains, and, more importantly, that contain a large number of spatial modes~\cite{Boyer2018Science}. We further take advantage of the ability of the FWM process to generate bright twin beams that contain a large number of correlated photons ($\sim$~mW per beam). This enhances the signal-to-noise ratio of the performed measurements to obtain a robust characterization of the spatial properties of the generated twin beams.

In the FWM process, two input pump photons are simultaneously converted into correlated probe and conjugate photon pairs (see Fig.~\ref{fig:setup}, top inset). As a result of phase-matching, the momentum correlations between the generated probe and conjugate are dictated by the momentum distribution, or angular spectrum, of the pump fields. In our experiments, the FWM is based on a configuration in which the two required pump photons come from a single pump field, as shown in Fig.~\ref{fig:setup}, which leads to the phase matching condition shown in the lower inset of Fig.~\ref{fig:setup} for the case of a plane-wave pump ($\delta$-function angular spectrum corresponding to a single $k$-vector). This leads to probe and conjugate photons with anti-correlated transverse emission directions with respect to the propagation direction of the pump, such that their transverse $k$-vectors are oriented in opposite directions, as shown in Fig.~\ref{fig:setup}. The momentum distribution, or $k$-vector content, of the twin beams can be mapped to a position distribution in the far field, which for a pump with a $\delta$-function angular spectrum translates to point-to-point correlations.  Any broadening of the pump angular spectrum, as  result of a localized pump beam (e.g. a Gaussian beam), would lead to a spread of the position correlations between the twin beams in the far field from point-to-point to point-to-area. We have experimentally verified such a distribution both in the near and far fields and shown that the spatial correlations are quantum in nature~\cite{ashok2017PRA,ashok2021IOP}. Going beyond such a spread requires additional modifications of the angular spectrum of the pump to further rearrange the distribution of the spatial correlations between the probe and conjugate.

The ability to modify the distribution of the spatial correlations in a controlled way, as needed to encode information, requires knowledge of the relationship between the pump's angular spectrum and the spatial correlations between the probe and conjugate. In particular, we consider the spatial cross-correlation between the twin beams, as its distribution in the far field gives information about the relative momentum distribution of photons in one of the beam conditioned upon a measurement of photons in the other beam with a specific momentum value. Furthermore, for the bright twin beams used in our experiments, previous studies have shown that the spatial quantum properties can be effectively characterized through a statistical analysis of the spatial intensity fluctuations of images of the twin beams acquired with a CCD~\cite{ashok2017PRA,Kumar18,ashok2019PRA,ashok2021IOP,lantz2010PRA}. In the bright limit the spatial cross-correlation, $c(\vec\xi)$, of the photon number fluctuations is equivalent to the spatial cross-correlation of the quadrature fluctuations (see Appendix~\ref{Sect:Cross}) such that
\begin{align}{\label{eq2}}
c(\bm\xi)&\equiv \expval{\delta \hat N_{pr}(-\bm{x})\delta \hat N_c(\bm{x}+\bm\xi)} \overset{\text{bright limit}}{\propto} \expval{\delta \hat X_{pr}(-\bm{x})\delta \hat X_c(\bm{x}+\bm\xi)},
\end{align}
where $\delta\hat N$ is the photon number fluctuation operator and $\delta\hat{X}$ is the amplitude quadrature fluctuation operator. The subscripts $pr$ and $c$ represent probe and conjugate fields, respectively, $\bm{x}$  is the two-dimensional transverse positions in the far field with its origin at the center of the pump beam, and $\bm\xi$ is the two-dimensional transverse position displacement.  Note that $c(\bm\xi)$ is translationally invariant with respect to $\bm{x}$.  The negative sign in the spatial argument of $\delta \hat N_{pr}$ results from the fact that correlated probe and conjugate are generated with opposite momenta in the transverse plane (following the phase matching condition).

To find the functional dependence of the spatial cross-correlation on the angular spectrum of the pump, we perform a perturbative expansion of the twin beam wavefunction to first order in the interaction Hamiltonian of the FWM. In this limit, we can show (see Appendix~\ref{Sect:Cross}) that the spatial cross-correlations takes the form
\begin{equation}{\label{convolution_relation}}
c(\bm\xi)\overset{\text{bright limit}}{\propto}\langle{\delta \hat X_{pr}(-\bm{x})\delta \hat X_c(\bm{x}+\vec{\xi})}\rangle \propto \mathfrak{Re}\left\{\Phi(\bm\xi)\right\},
\end{equation}
where $\Phi = (\mathcal{E}_o\star\mathcal{E}_o)$. Here, $\mathcal{E}_o$ is the angular spectrum of the pump,  which is given by the Fourier transform of the transverse pump field at the center of the nonlinear medium, and $\star$ denotes the convolution operation. Equation~(\ref{convolution_relation}) shows that to first order the distribution of the far-field spatial correlations is determined by the convolution of the angular spectra of the two pump photons that participate in the FWM process.

We take advantage of Eq.~(\ref{convolution_relation}) to encode information in the distribution of the spatial correlations of twin beams by implementing the FWM with a spatially structured high power laser beam as the pump (see Appendix~\ref{Sect:Methods}) in a $^{85}$Rb vapor cell. Bright twin beams are generated by seeding the FWM process with a weak probe beam (see Fig.~\ref{fig:setup} and Appendix~\ref{Sect:Methods}). A $f$-to-$f$ optical system is then used to map the momentum distribution of the twin beams to a position distribution in the far field that is imaged by an EMCCD, as shown in Fig.~\ref{fig:setup}. In order to extract the spatial intensity fluctuations to calculate the spatial cross-correlation, and thus read out the encoded information, two images are taken in rapid succession with the EMCCD and subtracted (see Appendices~\ref{Sect:Methods} and~\ref{Sect:data}). This approach also cancels any classical noise introduced by the seed beam~\cite{ashok2019PRA}.

\begin{figure}[htb]
	\centering
	\includegraphics{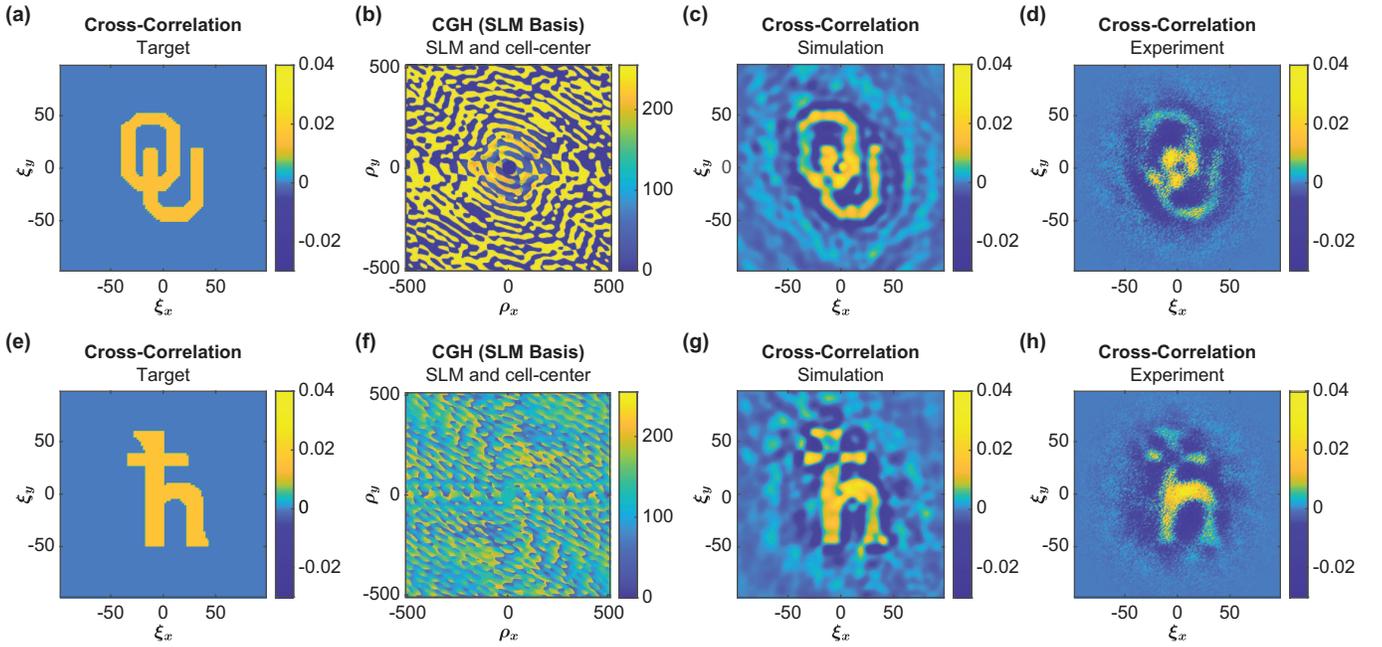}
	\caption{Information encoding in the distribution of the spatial correlations of twin beams. Figures (a) and (e) show the target information to be encoded in the spatial cross-correlation of the twin beams. The target is used to calculate the corresponding CGH (b) and (f) with a MRAF algorithm. The dimensions of the SLM pixels (12.5~$\mu$m$\times$12.5~$\mu$m) and its 8-bit resolution together with the $f$-to-$f$ imaging system are taken into account to calculate the simulated cross-correlations in frames (c) and (g). The measured spatial cross-correlations between the probe and conjugate beam intensity fluctuations reveal the encoded information, as shown in frames (d) and (h). Except for frames (b) and (f), each pixel value is normalized to the sum of the amplitude squared of all the pixels in the image to provide a better comparison between the simulation and experiment. The maximum values for the cross-correlations of the experimental and simulated data are larger than for the target due to the non-uniform distributions that result from a non-ideal setup and CGH. One can notice a small rotation ($\sim$ 4$^\circ$) in the measured spatial cross-correlations, which is due to experimental alignment imperfections. All figures, except for the CGH, are in the EMCCD pixel basis, with a pixel size of 16~$\mu$m$\times$16~$\mu$m. The color bar for the CGH frames (b) and (f) correspond to the 8-bit encoding of the phase in the range of 0 to $2\pi$. For a detailed explanation of the measurement procedure and calculation of the spatial cross-correlations, see Appendices~\ref{Sect:Methods} and~\ref{Sect:data}.}	\label{Fig2}
\end{figure}

To illustrate that is possible to encode information in the spatial correlations, we choose two patterns, the University of Oklahoma logo (OU) and {\textcrh} (Planck's constant), to be encoded in the twin beams as target far-field spatial correlation distributions (see Fig.~\ref{Fig2}). Here, we restrict to only phase changes to the pump beam to achieve the required angular spectrum. For each target, we imprint the necessary phase distribution on the pump beam with a phase-only computer generated hologram (CGH) implemented with a spatial light modulator (SLM), as shown in Fig.~\ref{fig:setup}. A $4f$-imaging system is used to perform a one-to-one imaging of the phase-structured pump reflected from the SLM to the cell center.  Given that the angular spectrum of the pump is a complex quantity, the implemented phase only CGH has to be designed to produce the calculated amplitude and phase for the pump's angular spectrum. Any deviation will affect the real part of the convolution function and hence the fidelity of the information encoded in the far-field spatial correlations. To minimize such deviations, we implement a mixed-region-amplitude-freedom (MRAF) algorithm coupled with  conjugate-gradient optimization to calculate the required CGH~\cite{Bowman2017OpEx} (see Appendix~\ref{Sect:CGH}).

Figure~\ref{Fig2}, which gives a comparison between target, simulated, and experimentally obtained far-field spatial cross-correlation patterns, shows the capability of our system to encode information in the distribution of the spatial correlations of the twin beams. The simulated patterns are based on Eq.~(\ref{convolution_relation}) and the corresponding calculated CGH, which takes into account the limited spatial and phase resolution (8-bit) of the SLM. As can be seen, there is excellent agreement between our experimental results and the expected spatial cross-correlations patterns obtained from the simulation. These results show that it is possible to achieve arbitrary spatial distributions in the twin beam correlations and thus represent a significant advancement with respect to previous proof-of-principle experiments with PDC~\cite{monken1998PRA,gigan2021OptL}. The fundamental limit to the resolution of the patterns that can be encoded is given by the number of spatial modes that the FWM process can support, which is mainly limited by the size of the pump beam as the use of an atomic system to implement the FWM makes it such that a cavity is not required to obtain a large nonlinear response. For our experiments, the pump size is limited by the maximum power output of the Ti:Sapphire laser. It is important to note that the main deviation from the target distribution comes from a non-optimal phase-only CGH to implement the required angular spectrum of the pump. This can be seen from the degradation of the resolution of the simulated spatial cross-correlation patterns in comparison to the target. There are several factors that lead to discrepancies between the simulated and measured cross-correlations. One is related to a decrease in efficiency of the SLM at higher spatial frequencies, which limits the spatial bandwidth of the angular spectrum imparted on the pump. A second one is that Eq.~(\ref{convolution_relation}) accounts only for the leading order contribution of the FWM interaction. Finally, the simulated patterns do not take into account the finite size of the bright probe and conjugate beams that effectively set the measurement region.

\begin{figure}[htb]
	\centering
	\includegraphics{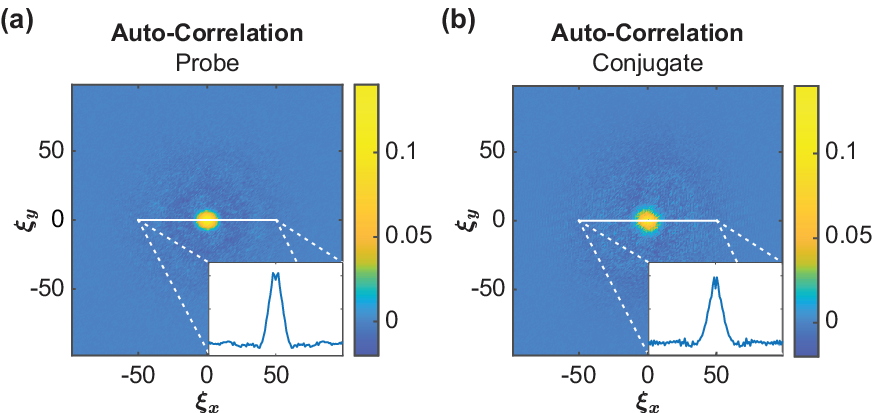}
	\caption{Experimental measurement of the auto-correlations of the spatial intensity fluctuations of the (a) probe and (b) conjugate fields. For these measurements the phase-structured pump beam is set to generate the OU logo pattern in the spatial cross-correlation between the twin beams, as shown in the top row of Fig.~\ref{Fig2}. While the encoded information can clearly be read out through joint measurements of the probe and the conjugate, each beam by itself (as seen by the auto-correlations) does not reveal the encoded information. The insets show a cross-section of the auto-correlations.  The measured auto-correlations are localized and almost identical to those obtained when the pump has not been modified (see Appendix~\ref{Sect:Auto}). An artificial peak at the center of the auto-correlations, which is due to the use of the same image to calculate them, is removed.}
	\label{auto_corr_main}
\end{figure}

For the distribution of the spatial correlations to be a viable degree of freedom for the secure transfer of information compatible with current approaches in quantum information science, it is necessary to preserve the temporal quantum correlations for security verification and for the encoded information to only be accessible through joint measurements of the twin beams. Since a modification of the angular spectrum of the pump only impacts the phase matching condition, its main effect is expected to be on the spatial properties of the generated twin beams and not the temporal ones. We have verified that the degree of intensity difference squeezing in the time domain is preserved when the pump is modified, with the main impact coming from the additional noise due to scattered pump photons (see Fig.~\ref{temporal_squeezing} in  the Appendix).

In terms of the spatial degree of freedom, each beam individually does not reveal the encoded information. While in principle the bright spatial profiles of the generated twin beams should be modified with a change of the angular spectrum of the pump, we avoid encoding information on the bright spatial profiles by adding a small constant-phase circular region at the center of the CGH, as can be seen in Figs.~\ref{Fig2}(b) and (f). The radius of this constant-phase region is kept as small as possible to minimize any impact on the CGH and thus the desired angular spectrum of the pump. The FWM process is aligned such that the seed probe beam overlaps with this constant phase region, which results in the generation of bright twin beams with Gaussian spatial profiles (see Fig.~\ref{supp_Fig2} in the Appendix). Additionally, as can be seen in Fig.~\ref{auto_corr_main}, the auto-correlation of the spatial fluctuations for each beam is localized and does not contain any of the encoded information. This is to be expected when the twin beams are composed of a large number of spatial modes that contribute to their spatial correlations (see Appendix~\ref{Sect:Auto}). Thus, the information encoded through the modification of the angular spectrum of the pump can only be recovered through joint measurements, such as the cross-correlation of the spatial fluctuations of the twin beams.

For a spontaneous (non-seeded) FWM process, the probe and conjugate fields are generated over a region centered around the pump with an angular bandwidth limited by the phase matching condition. The phase matching condition determines the spatial eigenmodes of the FWM process into which probe and conjugate photons are emitted. When seeded, in addition to these spontaneously generated photons, the FWM generates bright probe and conjugate beams that are localized to a subregion of the spontaneous emission region, as determined by the size of the seed probe beam and its angle with respect to the pump. The bright portions of the twin beams effectively act as local oscillators (LOs), one for the probe and one for the conjugate, for the FWM eigenmodes they overlap with, which leads to the amplification of the spatial fluctuations. This is analogous to measurements of quantum properties of bright twin beams in the time domain, where the bright portion serves as a LO to amplify the temporal fluctuations. As a result, for the measurements performed with the EMCCD, the bright twin beams provide an effective self homodyning of the FWM eigenmodes whose spatial profiles overlap with the Gaussian spatial profiles of the bright probe and conjugate beams. When the angular spectrum of the pump is modified, the spatial eigenmodes of the FWM are correspondingly modified, which leads to the changes in the spatial cross-correlation.

\begin{figure}
	\centering
	\includegraphics{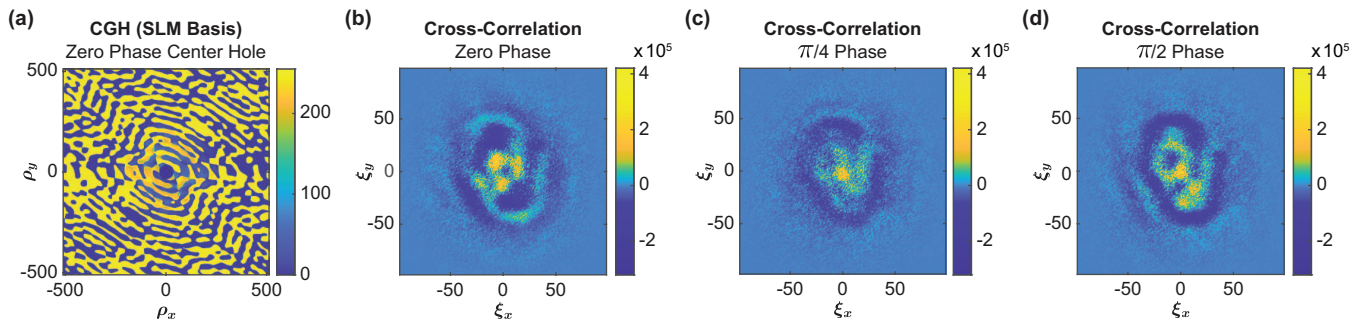}
	\caption{Control of the degree of correlation between the twin beams through the use of a CGH with a constant-phase central region. (a) OU logo CGH with zero-phase circular region at the center. Choosing a suitable change in phase ($\Delta\phi_{p}$)  for the constant-phase region without altering the rest of the CGH makes it possible to change the degree of correlation from (b) positively correlated ($\Delta\phi_{p}=0$) to (c) uncorrelated ($\Delta\phi_{p}=\pi/4$) to (d) negatively correlated ($\Delta\phi_{p}=\pi/2$). Figures (b), (c), and (d) are not normalized to obtain an absolute comparison of the degree of correlation as the phase of the constant-phase central region is changed.}
	\label{fig:Fig3}
\end{figure}

As described above, we add a constant-phase circular region at the center of the phase-structured pump to maintain the Gaussian spatial profiles of the bright probe and conjugate beams. An additional advantage of this approach is that it allows us to control the relative phase between the bright portions of the probe and conjugate, which act as LOs. In particular, for the FWM process in which two photons are absorbed from a single pump beam, the involved fields have to satisfy the phase relation
\begin{equation}{\label{phase_relation}}
2\phi_{p} = \phi_{pr} + \phi_{c},
\end{equation}
where the subscript $p$, $pr$, and $c$ stand for the pump, probe, and conjugate fields, respectively, and $\phi$ denotes the phase of the respective field. This relation leads, for example, to optical phase-conjugation~\cite{boyd2008book} for the generated conjugate for the case in which $\phi_{p}=0$. In our experiments, the phase-only CGH is generated with the constraint of having a constant-phase circular region at the center. This leads to an effective self homodyning of the probe and the conjugate for which the corresponding amplitude quadratures are naturally measured, as given by Eq.~(\ref{eq2}). If we now locally change the phase value of the constant-phase circular region by $\Delta\phi_{p}$ while keeping the phase of the rest of the CGH intact, we can change the phase of the generated bright conjugate with respect to the bright probe by $\phi_{c}=2\Delta\phi_{p}$, as dictated by Eq.~(\ref{phase_relation}), without a significant modification of the spatial eigenmodes of the FWM process.  This effectively rotates the phase of the conjugate LO with respect to the generated spatial eigenmodes, which makes it possible to select the quadrature that is being measured for the conjugate beam.
Figure~\ref{fig:Fig3} shows the measured spatial cross-correlation for $\Delta\phi_{p}=0$, $\Delta\phi_{p}=\pi/4$, and $\Delta\phi_{p}=\pi/2$. As expected, the correlations are significantly reduced for $\Delta\phi_{p}=\pi/4$, as ideally this corresponds to a measurement of uncorrelated quadratures in the form of  $\expval{\delta \hat X_{pr}(-\bm{x})\delta \hat Y_c(\bm{x}+\bm\xi)}$, and become negative for $\Delta\phi_{p}=\pi/2$, as the measurement is proportional to $\expval{\delta \hat X_{pr}(-\bm{x})\left(-\delta \hat X_c(\bm{x}+\bm\xi)\right)}$. This process can be used in general for any given CGH to further control the degree and sign of the correlations.

In summary, we have shown that it is possible to encode information in the distribution of the spatial correlations of twin beams. This is achieved through a complete control of the spatial correlations in the bright twin beams generated using FWM with a phase-structured pump beam. The degree of control we are able to obtain is key to enable information encoding in the form of a target spatial distribution and is a significant advancement over previous experiments with PDC~\cite{monken1998PRA,gigan2021OptL}. Furthermore, the information cannot be read out from either of the individual beam and the additional use of a constant-phase region at the center of the CGH prevents the information from being present in the bright spatial profiles of the generated twin beams.  Therefore, the capability to engineer the distribution of the spatial correlations directly at the source provides us with a novel approach to explore the use of the spatial structure of entangled bright twin beams for applications in quantum information science and could enable a source with on-demand entanglement between any arbitrary superposition of spatial modes.

Given that optical quantum correlations, which can be present in both the temporal and spatial degrees of freedom, form the basis for many emerging quantum technologies, the results presented here promise to have a significant impact in quantum information science. For example, the ability to encode information in the spatial degree of freedom of twin beams, such that it is only accessible through joint measurements, while simultaneously preserving the temporal quantum correlations opens the possibility of using spatial correlations for high capacity secure exchange of quantum information that takes advantage of the infinite-dimensional Hilbert space of spatial modes~\cite{forbes2020OSA}. On the other hand, for quantum sensing applications, a tunable multi-spatial mode quantum state can lead to enhanced imaging and positioning via the tailored distribution of spatial modes~\cite{kolobov2007book}. Furthermore, the bright nature of the generated multi-spatial mode states is particularly useful for applications demanding real-time imaging based on single/few-shot measurements~\cite{ashok2021IOP}. From a fundamental perspective, control of spatial correlations directly at the source provides a way to systematically study and understand the limits on the control of quantum correlations between various spatial modes. This knowledge could be beneficial for exploring potential applications in quantum information and scalable quantum computation~\cite{tasca2011PRA}.

\section*{Acknowledgments}
This work was supported by the National Science Foundation (NSF grant PHYS-1752938). A.M.M. acknowledges support from the US Department of Energy, Office of Science, National Quantum Information Science Research Centers, Quantum Science Center.

\appendix

\section{Methods}\label{Sect:Methods}

\noindent {\bf Experimental Details:} In order to generate quantum-correlated twin beams, referred to as probe and conjugate, we use a FWM process based on a double-$\Lambda$ system in a non-collinear configuration, as shown in Fig.~\ref{fig:setup}. During the FWM process, two pump photons from a single pump beam are simultaneously converted into a pair of probe and conjugate photons. Due to their simultaneous generation, the probe and conjugate beams are entangled in the temporal domain~\cite{Boyer2018Science}. When seeded, the FWM amplifies the input probe beam and generates a bright conjugate beam to produce bright twin beams. Tuning the number of photons in the input seed allows us to control the number of correlated photons in the generated bright twin beams. We implement the FWM in the D1 line of $^{85}$Rb in a vapor cell with a 1~inch diameter and a length of 12~mm. The pump beam has a power $\sim 2$ W and is locked with a sideband saturation spectroscopy lock 1.38~GHz to the blue of the $F=2\rightarrow F'=3$ transition. A high frequency acousto-optic modulator is used to red-shift a small portion of the pump by 3.04~GHz to generate the input seed probe. Orthogonally polarized pump and probe beams are made to intersect at an angle of 0.4 degrees at the center of the Rb vapor cell, which is held at a temperature of 114$^\circ$C. Optical systems are used before the cell to position the waists of both beams at the center of the cell with a $1/e^{2}$ waist diameter of 4.4~mm for the pump and 0.4 mm for the probe. With this configuration, we obtain a gain of $\sim 2.6$, which is limited by the power available for the pump beam. After the Rb vapor cell, a polarization filter separates the twin beams from the strong pump beam before they are imaged by the EMCCD.  While most of the pump is blocked with the polarization filter, the use of a structured pump significantly increases the amount of scattered pump light that makes it through the filter and onto the EMCCD. These uncorrelated photons significantly affect the cross-correlation measurements and can even saturate the EMCCD. In order to absorb the unwanted scattered pump photons selectively, we place an additional isotopically pure $^{87}$Rb 3-inch long vapor cell heated to 97$^\circ$C before the EMCCD camera. We set the detuning of the pump beam to maximize pump absorption while minimizing the losses for the probe and conjugate, which leads to a larger than optimal one-photon detuning for the FWM process. With this approach, we can obtain a pump transmission of $\sim 10$\% through the $^{87}$Rb cell, while the probe and conjugate transmission is kept at $\geq 90$\%.
\\

\noindent {\bf Correlation Measurements:} For the correlation measurements, a 500 mm lens is placed in a $f$-to-$f$ configuration between the center of the vapor cell and the EMCCD. The $f$-to-$f$ optical system maps the transverse momenta of the fields at the cell center to transverse position on the EMCCD (see Fig.~\ref{fig:setup}), such that a photon with transverse momentum ($\mathbf k^\perp$) is mapped to transverse position $\bm{x} = f \mathbf k^\perp/k$ in the far field, where $k$ is the magnitude of the total momentum of the photon. Given that the characterization of the distribution of the spatial correlation is based on calculating the cross-correlation of the spatial intensity fluctuations of the twin beams, as given by Eq.~(\ref{eq2}), we implement a detection scheme in which two frames are acquired in fast succession using the kinetics mode of the EMCCD. The kinetics mode allows us to capture multiple frames in a single acquisition from the EMCCD. Each frame consists of a probe and a conjugate image captured at the same time. Successive frames, and thus images for the probe and conjugate, can be taken with a time difference of $\sim 60~\mu$s between them~\cite{ashok2019PRA}. A subtraction of two successive frames provides images of the spatial intensity fluctuations in each beam, which are then use to evaluate the two-dimensional  spatial cross-correlation needed to characterize the distribution of the spatial correlations. Before calculating the cross-correlation, the image of the conjugate spatial fluctuations is rotated by $180^\circ$ to account for the transverse momentum anti-correlations between the probe and conjugate that result from phase matching. Additional details can be found in Appendix~\ref{Sect:data}.

\section{Spatial cross-correlation}\label{Sect:Cross}

\begin{figure}[h]
	\centering
	\includegraphics{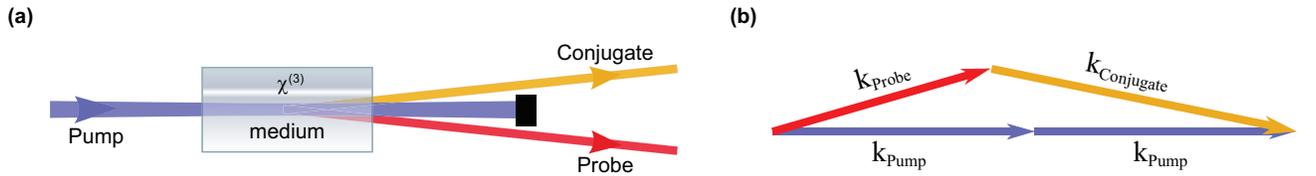}
	\caption{(a) Schematic diagram of the FWM process used in the experiment in which two photons are absorbed from a single pump field to generate quantum correlated probe and conjugate fields. (b) Phase matching condition for the configuration in which two pump photons are absorbed from a single pump field. Phase matching needs to be satisfied for an efficient FWM process and leads to momentum correlations between the generated probe and conjugate photons.  The non-collinear configuration results from the response of the atomic medium used for the FWM process~\cite{Turnbull2013}.}
	\label{ExFig2}
\end{figure}

In order to show the dependence of the spatial cross-correlation on the angular spectrum of the pump given by Eq.~(\ref{convolution_relation}) in the main text, we start with the interaction Hamiltonian for the four-wave mixing (FWM) process, which can be written as~\cite{slusher1987JOSAB},
\begin{equation}{\label{eq1}}
\mathcal{\hat H} = i \hslash\int {d}^{3}\mathbf{r}  \chi^{(3)}(\mathbf{r}) \hat E_{{p}}^{(+)}(\mathbf{r}, t) \hat E_{{p}}^{(+)}(\mathbf{r}, t) \hat{E}_{pr}^{(-)}(\mathbf{r}, t) \hat{E}_{c}^{(-)}(\mathbf{r}, t)+\mathrm{H.c.}.
\end{equation}
Here $\chi^{(3)}$ is the third order nonlinear response of the atomic media and the subscripts ${p}$, ${pr}$, and ${c}$ denote pump, probe, and conjugate fields, respectively. We have also assumed that the two required pump photons come from a single pump beam, as is the case for our experiment. These two pump photons are simultaneously converted into a pair of photons, one for the probe  and one for the conjugate, as shown in  Fig.~\ref{ExFig2}(a). In the undepleted pump approximation we can treat the pump as a classical field.  If we take the pump to be propagating along the $z$-direction, it can be written as
\begin{equation}{\label{eqS2}}
E_{p}^{+}(\mathbf r, t)= E_o(\bm \rho, z) e^{i\left(k^{z}_{p}z-\omega_p t\right)},
\end{equation}
where $k^{z}_{p}$ denotes the $z$-component of the pump's $k$-vector, $\bm\rho$ is the two-dimensional position in the transverse plane at the center of the nonlinear medium, and $E_o(\bm \rho, z)$ is a complex function representing the slowly varying envelop of the pump's electric field. In our experiments, the waist of the pump is placed at the center of the $\chi^{(3)}$ nonlinear medium (atomic vapor cell) and its transverse spatial distribution is assumed to remain unchanged throughout the length of the nonlinear medium. As a result, the slowly varying field amplitude is assumed to be independent of $z$, that is $E_o(\bm \rho,z)=E_o(\bm \rho)$.

In momentum space ($k$-space) the field operators for the probe and conjugate can be written as
\begin{eqnarray}
\hat E_{pr}^{-}(\mathbf r, t)&=& \left(\frac{1}{2\pi}\right)^{3/2} \int d \mathbf k_{pr}\,  e^{-i\left(\mathbf k_{pr}\cdot\mathbf r-\omega_{pr} t\right)} \hat a_{\mathbf k_{pr}}^{\dagger},\label{eq3}\\
\hat E_{c}^{-}(\mathbf r, t)&=&\left(\frac{1}{2\pi}\right)^{3/2} \int d \mathbf k_{c}\,  e^{-i\left(\mathbf k_{c}\cdot\mathbf r-\omega_c t\right)} \hat b_{\mathbf k_{c}}^{\dagger},\label{eq4}
\end{eqnarray}
where $\hat a_{\mathbf k_{pr}}^{\dagger}$ ($\hat b_{\mathbf k_{c}}^{\dagger}$) is the creation operator for a probe (conjugate) photon with a spatial profile given by a plane wave and with $\hslash \mathbf{k}$ momentum. Using Eqs.~(\ref{eqS2}), (\ref{eq3}), and (\ref{eq4}), the FWM interaction Hamiltonian can be rewritten as
\begin{align}
\mathcal{\hat H} &= i \hslash \left(\frac{1}{2\pi}\right)^{3}\int \int \int {d}^{3}\mathbf{r}\,  d \mathbf k_{pr}\,  d \mathbf k_{c}\, \chi^{(3)}(\mathbf{r})  E^2_o(\bm \rho) e^{2i\left(k^{z}_{p}z-\omega_p t\right)}  e^{-i\left(\mathbf k_{pr}\cdot\mathbf r-\omega_{pr} t\right)} e^{-i\left(\mathbf k_{c}\cdot\mathbf r-\omega_{c} t\right)} \hat a_{\mathbf k_{pr}}^{\dagger} \hat b_{\mathbf k_{c}}^{\dagger}+\mathrm{H.c.} {\label{eq5}}\\
& = i \hslash\Gamma \int \int \int {d}^{3}\mathbf{r}\,  d \mathbf k_{pr}\,  d \mathbf k_{c}\,   E^2_o(\bm \rho) e^{-i\left(\mathbf k^\perp_{pr} +\mathbf k^\perp_{c} \right)\cdot\bm \rho} e^{i\Delta k_z z } \hat a_{\mathbf k_{pr}}^{\dagger} \hat b_{\mathbf k_{c}}^{\dagger}+\mathrm{H.c.}, {\label{eq6}}
\end{align}
where we have assumed that $\chi^{(3)}$ is spatially independent so that it can be taken outside the integral and absorbed into $\Gamma = \chi^{(3)}/(2\pi)^{3}$. The term $\Delta k_z = 2k^{z}_{p} -k^{z}_{pr} - k^{z}_{c}$ denotes the longitudinal phase-mismatch , and $\mathbf k^{\perp}_{pr}$ ($\mathbf k^{\perp}_{c}$) is the probe (conjugate) momentum vector is the transverse plane. In obtaining Eq.~(\ref{eq6}) we have assumed the angular frequency mismatch ($\Delta \omega = 2\omega_p - \omega_{pr} -\omega_c$) to be zero.

In order to highlight the dependence of the FWM on the transverse spatial profile of the pump, we further write the interaction Hamiltonian as
\begin{align}
\mathcal{\hat H} & = i \hslash 2\pi\Gamma \int \int \int dz\, d \mathbf k_{pr}\,  d \mathbf k_{c}\,   \left[\frac{1}{2\pi}\int d\bm\rho\,  E^2_o(\bm \rho) e^{-i\left(\mathbf k^\perp_{pr} +\mathbf k^\perp_{c} \right)\cdot\bm \rho} \right] e^{i\Delta k_z z } \hat a_{\mathbf k_{pr}}^{\dagger} \hat b_{\mathbf k_{c}}^{\dagger}+\mathrm{H.c.} {\label{eq7}} \\
& = i \hslash 2\pi\Gamma \int \int \int dz\, d \mathbf k_{pr}\,  d \mathbf k_{c}\,  \mathbb K(\mathbf k^\perp_{pr} +\mathbf k^\perp_{c})  e^{i\Delta k_z z } \hat a_{\mathbf k_{pr}}^{\dagger} \hat b_{\mathbf k_{c}}^{\dagger}+\mathrm{H.c.},{\label{eq8}}
\end{align}
where we have introduced the function $ \mathbb K(\mathbf k^\perp_{pr} +\mathbf k^\perp_{c}) \equiv \frac{1}{2\pi}\int d\bm\rho\, E^2_o(\bm \rho) e^{-i\left(\mathbf k^\perp_{pr} +\mathbf k^\perp_{c}\right)\cdot\bm \rho}$, which contains all the information of the pump transverse spatial profile and represents the transverse Fourier transform of $E^2_o(\bm \rho)$.
We can further simply this expression by writing the pump's transverse spatial distribution in momentum space, such that
\begin{equation}{\label{eq9}}
E_o(\bm \rho)=\frac{1}{2\pi}\int d \mathbf k^\perp_{p}\, \mathcal E_o(\mathbf k^\perp_{p}) e^{i\mathbf k^\perp_{p } \cdot \bm\rho},
\end{equation}
where $\mathcal E_o(\mathbf k^\perp_{p})$ represents the angular spectrum of the pump.  We can then substitute Eq.~(\ref{eq9}) in the expression for $ \mathbb K$ to rewrite it as
\begin{align}
\mathbb K(\mathbf k^\perp_{pr} +\mathbf k^\perp_{c}) &= \frac{1}{(2\pi)^3}\int d\bm\rho\, \left(\int d \mathbf k^\perp_{p}\, \mathcal E_o(\mathbf k^\perp_{p }) e^{i\mathbf k^\perp_{p } \cdot \bm\rho} \int d \mathbf k'^\perp_{p}\, \mathcal E_o(\mathbf k'^\perp_{p }) e^{i\mathbf k'^\perp_{p } \cdot \bm\rho} \right) e^{-i\left(\mathbf k^\perp_{pr} +\mathbf k^\perp_{c}\right)\cdot\bm \rho} \\
&= \frac{1}{(2\pi)^3} \int \int d \mathbf k^\perp_{p}\, d \mathbf k'^\perp_{p }  \mathcal E_o(\mathbf k^\perp_{p }) \mathcal E_o(\mathbf k'^\perp_{p })\int d\bm\rho\, e^{i\left(\mathbf k^\perp_{p }+\mathbf k'^\perp_{p }-\mathbf k^\perp_{pr} -\mathbf k^\perp_{c}\right)\cdot\bm \rho} \\
&= \frac{1}{(2\pi)}\int \int d \mathbf k^\perp_{p}\, d \mathbf k'^\perp_{p}\, \mathcal E_o(\mathbf k^\perp_{p }) \mathcal E_o(\mathbf k'^\perp_{p }) \delta(\mathbf k'^\perp_{p }-(\mathbf k^\perp_{pr} +\mathbf k^\perp_{c} - \mathbf k^\perp_{p })) \label{eq12}\\
&= \frac{1}{(2\pi)}\int d \mathbf k^\perp_{p}\,  \mathcal E_o(\mathbf k^\perp_{p}) \mathcal E_o(\mathbf k^\perp_{pr} +\mathbf k^\perp_{c} - \mathbf k^\perp_{p }) \\
&= \frac{1}{(2\pi)}\Phi(\mathbf k^\perp_{pr} +\mathbf k^\perp_{c}), \label{eq15}
\end{align}
where $\Phi(\mathbf k^\perp_{pr} +\mathbf k^\perp_{c})\equiv\int d \mathbf k^\perp_{p}\,  \mathcal E_o(\mathbf k^\perp_{p }) \mathcal E_o(\mathbf k^\perp_{pr} +\mathbf k^\perp_{c} - \mathbf k^\perp_{p })$ is the convolution of the angular spectra of the two pump photons involved in the FWM process.
Finally, substituting Eq.~(\ref{eq15}) into Eq.~(\ref{eq8}) and integrating over the length $L$ of the media (from $-L/2$ to $L/2$) allows us to write the interaction Hamiltonian as
\begin{align}
\mathcal{\hat H} & = i \hslash \Gamma L \int \int  d \mathbf k_{pr}\,  d \mathbf k_{c}\,  \Phi(\mathbf k^\perp_{pr} +\mathbf k^\perp_{c}) \sinc(\Delta k_z L/2) \hat a_{\mathbf k_{pr}}^{\dagger}  \hat b_{\mathbf k_{c}}^{\dagger}+\mathrm{H.c.}{\label{pre_final_H}} \\
& = i \hslash \Gamma L \int \int  d \mathbf k_{pr}\,  d \mathbf k_{c}\,  \mathbb{F}(\mathbf k_{pr} ,\mathbf k_{c}) \hat a_{\mathbf k_{pr}}^{\dagger}  \hat b_{\mathbf k_{c}}^{\dagger} + \mathrm{H.c.},{\label{final_H}}
\end{align}
where we have defined the two-photon amplitude function $\mathbb{F}(\mathbf k_{pr} ,\mathbf k_{c}) \equiv \Phi(\mathbf k^\perp_{pr} +\mathbf k^\perp_{c}) \sinc(\Delta k_z L/2)$, which takes into account the angular spectrum of the pump through function $\Phi$  and the phase matching condition through the $\sinc$ function.
Note that the interaction Hamiltonian is very similar to the one for parametric down-conversion (PDC)~\cite{schneeloch2016IOPJOpt,barbosa2000PRL,shih1996PRA}, except  that in PDC the function $\Phi$ is directly given by the angular spectrum of the single pump photon involved in the process instead of the convolution of the angular spectra of the two pump photons involved in FWM.

With this interaction Hamiltonian we can approximate the state of the generated twin beams by using first order perturbation theory to write the twin beam state (TBS) wavefunction as
\begin{align}
\ket{\Psi_{TBS}} & = e^{-\frac{i\mathcal{\hat H}t}{\hslash}} \ket{\Psi_0} \simeq\left(1-\frac{i\mathcal{\hat H} t}{\hslash }\right) \ket{\Psi_0} \\
& =\ket{\Psi_0} + C_1 \int \int  d \mathbf k_{pr}\,  d \mathbf k_{c}\,  \left[ \mathbb{F}(\mathbf k_{pr} ,\mathbf k_{c}) \hat a_{\mathbf k_{pr}}^{\dagger}  \hat b_{\mathbf k_{c}}^{\dagger} + \mathbb{F}^{*}(\mathbf k_{pr} ,\mathbf k_{c}) \hat a^{ }_{\mathbf k_{pr}}  \hat b_{\mathbf k_{c}} \right]\ket{\Psi_0} \\
& = \ket{\Psi_0} + C_1 \int \int  d \mathbf k_{pr}\,  d \mathbf k_{c}\,  \mathbb{F}(\mathbf k_{pr} ,\mathbf k_{c}) \hat a_{\mathbf k_{pr}}^{\dagger}  \hat b_{\mathbf k_{c}}^{\dagger} \ket{\Psi_0} ,{\label{TBS}}
\end{align}
where we have assumed the input state to be the multimode vacuum state $\ket{\Psi_0} = \ket{\{0\}_{\mathbf k_{pr}},\{0\}_{\mathbf k_{c}}}$ with notation $\ket{\{0\}_{\mathbf k}}=\prod_{\mathbf k}\ket{0_{\mathbf k}}$ and $C_1 = \Gamma L \,t$ with interaction time $t$. It is important to note that even though we seed the process for the present experiment, the seed does not have an impact on the eigenmodes of the FWM and only serves to generate bright probe and conjugate beams with Gaussian profiles that serve as local oscillators for measuring the correlations in the FWM eigenmodes, as described in the main text.
The form of the TBS wavefunction given in Eq.~(\ref{TBS}) is the same as the one from type-I SPDC~\cite{schneeloch2016IOPJOpt,barbosa2000PRL,monken2004PRA,shih1996PRA,barbosa2005PRL,arnaut2002PRA} with the angular spectrum of pump field replaced with the convolution of the angular spectra of the two pump photons involved in the FWM process. The second term in Eq.~(\ref{TBS}) is responsible for the momentum correlations between the probe and conjugate beams with a distribution dictated by $\mathbb{F}(\mathbf k_{pr} ,\mathbf k_{c}$), such that if a single $k$-vector for the probe is measured the distribution of the corresponding correlated $k$-vectors for the conjugate can be engineered by  changing the angular spectrum of the pump.

With the TBS wavefunction given in Eq.~(\ref{TBS}), we are now in a position to calculate the spatial cross-correlation defined in the main text in terms of the spatial intensity fluctuations of images acquired by the EMCCD. We start by noting that a $f$-to-$f$ optical system is used to map the momentum distribution to a position distribution on the EMCCD in the far field; hence, the transverse spatial cross-correlation of the spatial intensity fluctuations in the far field can be written as
\begin{equation}
	c(\bm{x}_1,\bm{x}_2)\equiv\expval{\delta \hat N_{pr}(\bm{x}_1;z_{f})\delta \hat N_c(\bm{x}_2;z_{f})}
	\overset{f-{\rm to}-f}{\underset{\rm{optical~system}}{=}}
	\expval{\delta \hat N_{pr}(\mathbf{k}^\perp_1;z_o)\delta \hat N_c(\mathbf{k}^\perp_2;z_o)}
    \equiv \left. \expval{\delta \hat N_{pr}(\mathbf{k}^\perp_1,k_{1}^{z})\delta \hat N_c(\mathbf{k}^\perp_2,k_{2}^{z})}\right|_{z=z_{0}}, \label{k_space_transform}
\end{equation}
where $z_{f}$ and $z_o$ indicate the far field and near field along the propagation direction ($z$-axis), respectively, $\delta\hat N = \hat N - \expval{\hat{N}}$ with photon number operator $\hat{N}$, and $\bm x_1$ and $\bm x_2$ are two-dimensional transverse positions in the far field. The measurements are done in the far field where a photon with a transverse momentum vector $\mathbf k^{\perp}$ is mapped to spatial location $\bm x=f \mathbf k^\perp/k$ by the $f$-to-$f$ optical system. In the last expression, the $z$-component of the $k$-vector is given by $k_{j}^{z}=\sqrt{k_{j}^2-|\mathbf{k}^\perp_j|^2}$, where $k_{j}$ is the magnitude of the corresponding $k$-vector.

We also note that in the bright limit the photon number fluctuations are proportional to the amplitude quadrature fluctuations of the field. This can be shown by writing the field operators as $\hat{a}=\abs{\alpha}+\delta\hat{a}$, with $\abs{\alpha}$ representing the mean value (amplitude) of the field and $\delta \hat a$ the field fluctuation operator, such that
\begin{equation}
\hat{N}= \hat a^\dagger \hat a = (\abs{\alpha} + \delta \hat a)^\dagger(\abs{\alpha} + \delta \hat a)  \simeq \abs{\alpha}^2 + \abs{\alpha} \left(\delta \hat a^\dagger+ \delta \hat a\right), \label{eq23}
\end{equation}
where we have taken advantage of the fact that in the bright limit $\abs{\delta\hat{a}}/\abs{\alpha}\ll 1$ to drop the term quadratic in the field fluctuations. Here we can identify $\expval{\hat{N}} = \abs{\alpha}^2$ and $\delta\hat{X}= (\delta\hat{a}^\dagger+\delta\hat{a})/\sqrt{2}$
to rewrite Eq.~(\ref{eq23}) as
\begin{equation}
\delta \hat N \simeq \sqrt{2} \abs{\alpha}\delta \hat X,
\end{equation}
which makes it possible to write the spatial cross-correlation in terms of quadrature operators
\begin{eqnarray}{\label{brightlimit}}
	c(\bm{x}_1,\bm{x}_2)&=& \expval{\delta \hat N_{pr}(\bm{x}_1;z_{f})\delta \hat N_c(\bm{x}_2;z_{f})}\\
	&\overset{\mathrm{bright~limit}}{\propto}& \expval{\delta \hat X_{pr}(\bm{x}_1;z_{f})\delta \hat X_c(\bm{x}_2;z_{f})}\\
	&\overset{f-{\rm to}-f}{\underset{\rm{optical~system}}{=}}&\expval{\delta \hat X_{pr}(\mathbf{k}^\perp_1;z_o)\delta \hat X_c(\mathbf{k}^\perp_2;z_o)}
\equiv \left. \expval{\delta \hat X_{pr}(\mathbf{k}_1)\delta \hat X_c(\mathbf{k}_2)}\right|_{z=z_{0}}\label{expectation_A}.
\end{eqnarray}
We can now take the first order approximation of the TBS wavefunction given in Eq.~(\ref{TBS}) to calculate the required expectation value of the spatial cross-correlation of the quadrature fluctuations in $k$-space given in Eq.~(\ref{expectation_A}). If we take into account that $\expval{\hat X_{\{pr,c\}}}$ vanishes for $\ket{\Psi_{TBS}}$, we have that
\begin{flalign}
\expval{\delta \hat X_{pr}(\mathbf{k}_1)\delta \hat X_c(\mathbf{k}_2)}		
&=\expval{\hat X_{pr}(\mathbf{k}_1)\hat X_c(\mathbf{k}_2)}\label{expectation_B} \\
&\simeq \expval{\Psi_{0}|\hat X_{pr}(\mathbf{k}_1)\hat X_c(\mathbf{k}_2)|\Psi_{0}} +2 \int \int  d \mathbf k_{pr}\,  d \mathbf k_{c}\, \mathfrak{Re}\left\{ C_1\mathbb{F}(\mathbf k_{pr} ,\mathbf k_{c}) \langle\Psi_0| \hat X_{pr}(\mathbf{k}_1)\hat X_c(\mathbf{k}_2)\hat a_{\mathbf k_{pr}}^{\dagger}\hat b_{\mathbf k_{c}}^{\dagger} \ket{\Psi_0} \right\} \nonumber \\
& \quad+ \abs{C_1}^2 \int \int  d \mathbf k_{pr} d \mathbf k_{c} \int \int  d \mathbf k'_{pr} d \mathbf k'_{c} \mathbb{F}(\mathbf k_{pr} ,\mathbf k_{c}) \mathbb{F}^{*}(\mathbf k'_{pr} ,\mathbf k'_{c})  \langle\Psi_0|\hat b_{\mathbf k'_{c}}\hat a_{\mathbf k'_{pr}}\hat X_{pr}(\mathbf{k}_1)\hat X_c(\mathbf{k}_2)\hat a_{\mathbf k_{pr}}^{\dagger}\hat b_{\mathbf k_{c}}^{\dagger} \ket{\Psi_0} \label{long_expectation_expression}\\
&= \int \int  d \mathbf k_{pr}\, d \mathbf k_{c}\mathfrak{Re}\left\{ C_1\mathbb{F}(\mathbf k_{pr} ,\mathbf k_{c}) \langle\Psi_0| (\hat a_{\mathbf k^{ }_{1}}\hat b^{ }_{\mathbf k_{2}} + \hat a^{ }_{\mathbf k_{1}}\hat b^\dagger_{\mathbf k_{2}} + \hat a^\dagger_{\mathbf k_{1}}\hat b^{ }_{\mathbf k_{2}} + \hat a^\dagger_{\mathbf k_{1}}\hat b^\dagger_{\mathbf k_{2}})\hat a_{\mathbf k_{pr}}^{\dagger}  \hat b_{\mathbf k_{c}}^{\dagger} \ket{\Psi_0}\right\}  \\
&= \int \int  d \mathbf k_{pr}\,  d \mathbf k_{c}\, \mathfrak{Re}\left\{ C_1\Phi(\mathbf k^\perp_{pr} +\mathbf k^\perp_{c}) \sinc(\Delta k_z L/2) \right\} \delta(\mathbf{k}_1-\mathbf{k}_{pr})
\delta(\mathbf{k}_2-\mathbf{k}_{c}) \\
&= \sinc((2k_{p}^{z} -k_{1}^{z} - k_{2}^{z})L/2) \mathfrak{Re} \left\{C_1\Phi(\mathbf k^\perp_{1} +\mathbf k^\perp_{2})\right\}. \label{final_expectation1}
\end{flalign}
For the case in which the correlations are measured within a small region around the optimal direction of the FWM process, the phase-mismatch ($\Delta k_z$) is close to zero and the sinc function can be taken to be unity~\cite{boyd1985OpticsComm,boyd1981PRL}. In our experiments, the regions in the far field that contribute to the measurements are selected by the spatial extent of the bright probe and conjugate beams and represent a small region of the full spatial bandwidth of the FWM process, as required to approximate the sinc function as a constant. Finally, taking into account the relation between position and momentum in the far field given by Eq.~(\ref{expectation_A}), we can express Eq.~(\ref{final_expectation1}) in position coordinates ($\bm x_1,~\bm x_2$) such that
\begin{equation}
c(\bm{x}_1,\bm{x}_2)
\overset{\mathrm{bright ~ limit}}{\propto} \expval{\delta \hat X_{pr}(\bm{x}_1;z_{f})\delta \hat X_c(\bm{x}_2;z_{f})}\propto \mathfrak{Re} \left\{\Phi(\bm x_{1} +\bm x_{2})\right\}, \label{final_expectation2}
\end{equation}
where we have taken constant $C_{1}$ to be real without loss in generality, as the phase of the nonlinear response of the atomic medium, given by $\chi^{(3)}$, can be set to zero and serve as a phase reference for the rest of the fields involved in the FWM process. Additionally, we note that as a result of the conservation of momentum, non-zero correlations are only possible if ${-\bm x_1} \approx {\bm x_2}$ (up to the uncertainty in transverse momentum). Thus, one can define $\bm\xi=\bm x_{1} +\bm x_{2}$ and rewrite Eq.~(\ref{final_expectation2}) as
\begin{align}
c(\bm{\xi})
\overset{\mathrm{bright ~ limit}}{\propto}\expval{\delta \hat X_{pr}(-\bm{x})\delta \hat X_c(\bm{x}+\bm{\xi})} \propto \mathfrak{Re}\left\{\Phi(\bm\xi)\right\}\label{eq24},
\end{align}
which corresponds to Eq.~(\ref{convolution_relation}) in the main text.

\section{Image acquisition and measurement of spatial cross-correlation}\label{Sect:data}
\begin{figure}[htb]
	\centering
	\includegraphics{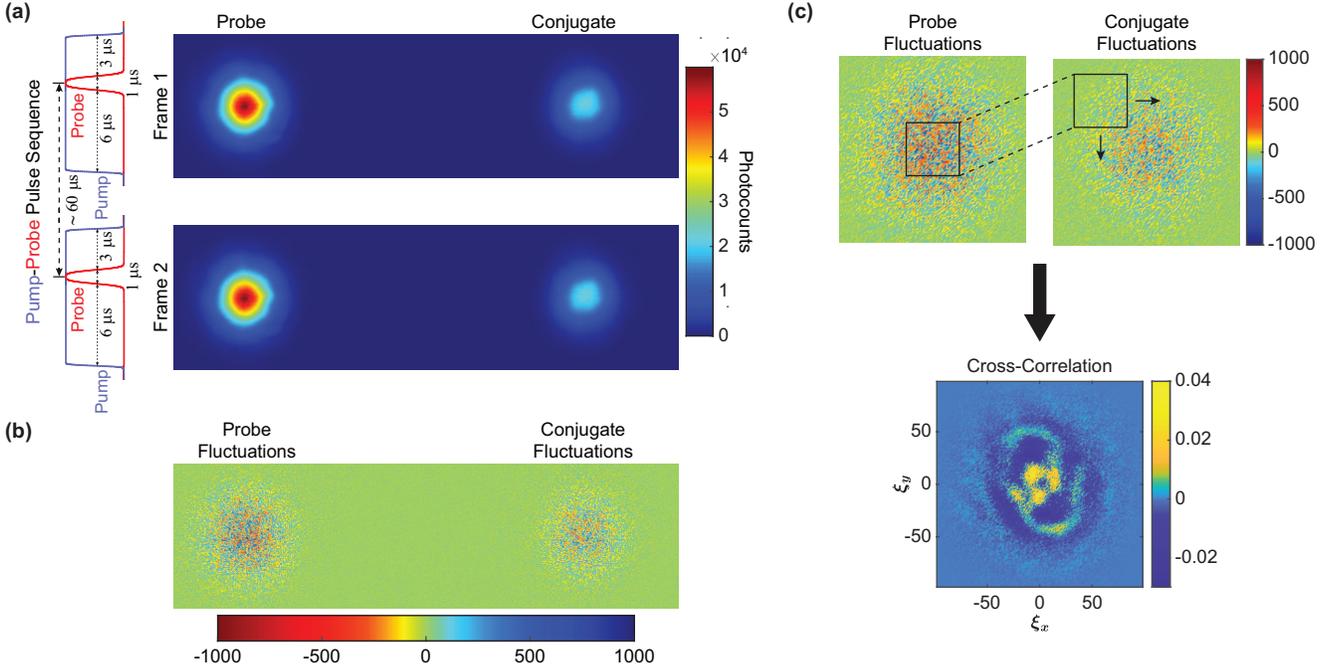}
	\caption{Image acquisition sequence and spatial cross-correlation of the spatial intensity fluctuations.  (a) Images of bright probe and conjugate beams in two consecutive frames. The pulse sequence, which is synchronized with the data acquisition of the EMCCD, is shown on the left.  The separation between two seed probe pulses is $\sim 60 \mu$s. (b) The spatial intensity fluctuation images for the probe (left) and conjugate (right) beams are obtained by performing a pixel to pixel subtraction of the two consecutive frames. (c) After rotation of one of the fluctuation images, a two dimensional spatial cross-correlation is implemented to calculate the distribution of the relative spatial correlations between the probe and the conjugate.}
	\label{supp_Fig2}
\end{figure}

In order to measure the spatial cross-correlation of the spatial intensity fluctuations, we acquire images of the bright probe and conjugate beams with an electron multiplying charge coupled device (EMCCD). To extract the spatial intensity fluctuations from these images, we take two frames (each with a bright probe and conjugate image) in rapid succession using the kinetics mode of the EMCCD~\cite{ashok2017PRA, ashok2019PRA}. The timing sequence for the pump and probe pulses for the two frames is shown in Fig.~\ref{supp_Fig2}(a). The input probe and pump beams are pulsed with a time duration of 1~$\mu$s and 10~$\mu$s, respectively, with the timing synchronized with the data acquisition by the EMCCD. The probe pulse is delayed by 6~$\mu$s with respect to the pump pulse to avoid transient effects in the FWM. The active area of the EMCCD is divided into frames with 170 (rows)~$\times$ 512~(columns) pixels.  Given the maximum charge transfer rate of 300~ns/row in the kinetics mode, the minimum time difference between two adjacent frames is  51~$\mu$s. The camera exposure time per frame is set to 12~$\mu$s, which leads to a time scale between two consecutive images of $\sim 60~\mu$s.

Figure~\ref{supp_Fig2}(a) shows bright probe and conjugate images acquired in two consecutive frames of the EMCCD. As can be seen, the peak region of the probe image has $\sim5\times10^4$ photocounts per pixel. The  probe and conjugate spatial intensity fluctuations, shown in Fig.~\ref{supp_Fig2}(b), are obtained by performing a pixel to pixel subtraction of the two consecutive frames. Such a differential analysis technique extracts the spatial fluctuations while reducing the common spatial classical noise present in the twin beams. Given that the two consecutive frames are taken with a time difference longer than the inverse of the bandwidth of the FWM process ($\sim$ 20 MHz), there are no quantum correlations between them, so the subtraction has no impact on the quantum properties of the twin beams.

To calculate the spatial cross-correlations of the spatial intensity fluctuations, we first rotate one of the fluctuation images by $180^\circ$ as required by the conservation of momentum, see Eq.~(\ref{eq24}). This allows us to obtain the translation invariant form of $\Phi$, and hence the spatial cross-correlation function $c(\bm\xi)$ by directly evaluating a two dimensional spatial cross-correlation between the probe and conjugate spatial intensity fluctuation images. As illustrated in top part of Fig.~\ref{supp_Fig2}(c), this is done by taking a portion of the spatial intensity fluctuations of the probe and using it to calculate the cross-correlation with the spatial intensity fluctuations of the conjugate as a function of the relative displacement, $\bm\xi$, between corresponding pixels. The process is repeated 2,000 times and the resulting spatial cross-correlations are then averaged to obtain the distribution shown in the bottom part of Fig.~\ref{supp_Fig2}(c).

\section{Implementation of computer generated hologram}\label{Sect:CGH}

\begin{figure}
	\centering
	\includegraphics{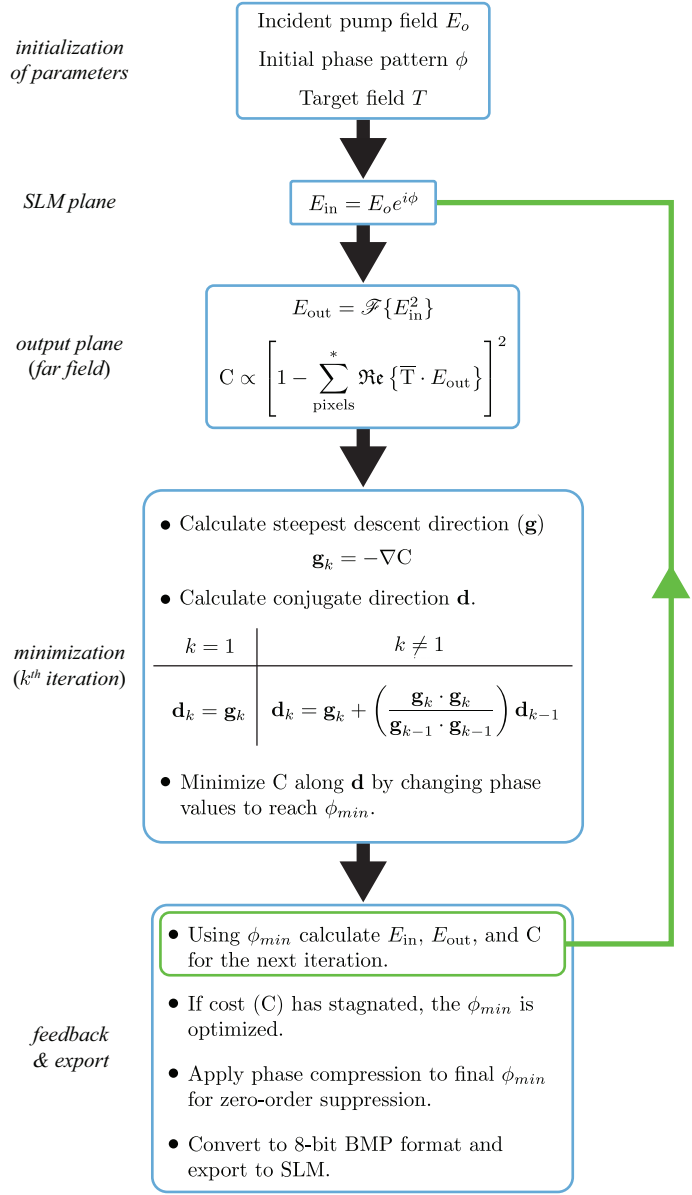}
	\caption{Calculation and optimization of the computer generated hologram. The flowchart shows the procedure for calculating the optimal $E_\text{out}$ using the conjugate minimization approach coupled with the MRAF algorithm. The pump beam is modified with a phase pattern at the SLM plane such that  $E_{\text{out}}$ in the far-field plane matches the desired target field $T$. The procedure is initialized with  an initial guess phase $\phi$ and the corresponding $E_\text{out}$ is compared with target pattern $T$ to estimate an initial cost using the cost function C. The cost function is then iteratively minimized along various conjugate directions ({\bf d}) until the cost function stagnates. After implementing a final phase-compression algorithm to eliminate the zero order from the SLM reflection, the final phase pattern is converted into BMP format and exported to the SLM.}
	\label{flowchart}
\end{figure}

As shown in the main text, we are able to engineer the distribution of the spatial correlations between the twin beams by using a structured pump beam. This requires imparting a specific angular spectrum on the pump, as dictated by Eq.~(\ref{eq24}). The desired pump structure, determined by $\Phi(\mathbf k^\perp_{pr} +\mathbf k^\perp_{c})$, is implemented via the numerical computation of a suitable phase pattern $\phi$ that is transferred to the pump beam with a spatial light modulator (SLM). The phase structured pump beam then goes through a $4f$-imaging system such that it is mapped to the center of the Rb vapor cell (pump beam waist location). A computer generated hologram (CGH) is used to impart the necessary phase distribution for a given target. The goal then is to calculate the necessary CGH with phase distribution $\phi(\bm \rho)$, with $\bm \rho = (\rho_1, \rho_2)$ the transverse position coordinate at the center of the cell, such that the pump field in the far field, $E_{\text{out}}$, matches the amplitude and phase of the target field distribution $T$, that is
\begin{align}
E_{\text{out}}  &= \mathcal{E}_o \left(\frac{f\mathbf k^\perp_p}{k}\right) \star \mathcal{E}_o \left(\frac{f\mathbf k^\perp_p}{k}\right) \\
&= \mathcal{F} \left\{\left[E_o(\bm \rho)e^{i\phi(\bm \rho)}\right]^2\right\}=T, \label{FFT_E_sq}
\end{align}
where $E_o(\bm \rho)$ is the pump field incident on the SLM, which for our case has a Gaussian profile and flat wavefront, and $\star$ denotes the convolution operation.

Figure~\ref{flowchart} shows the procedure to calculate and optimize the CGH to obtain the required angular spectrum for the pump.  The incident pump field on the SLM, $E_o$, is multiplied with an initial guess phase pattern $\phi$ to initialize the input field, $E_\text{in}$.  This initial field is then used as a starting point to calculate $E_\text{out}$ and subsequently used to optimize the phase distribution $\phi$ by assigning a cost to any deviations from the target pattern $T$ and using a minimization algorithm to reduce that cost. A low cost value gives a high degree of overlap between $E_{\text{out}}$ and $T$ and ensures an optimal angular spectrum for the pump field. For the optimization of $E_\text{out}$, we use a conjugate minimization algorithm~\cite{Bowman2017OpEx} coupled with the mixed-region-amplitude-freedom (MRAF) approach. In MRAF, the transverse plane of target field is divided into a signal and a noise region and the cost function is only evaluated over the signal region, thus allowing $E_{\text{out}}$ to take any values outside this region. As a result, a mismatch between $T$ and $E_{\text{out}}$ in the noise region does not affect the cost values.

For the minimization using the conjugate gradient minimization algorithm, we define the cost function (C) as
\begin{equation}{\label{cost function}}
\mathrm{C} = 10^d \left[1-\sum_{\text{pixels}}^{*} \mathfrak{Re}\left\{\overline{T} \cdot E_{\text{out}}\right\}\right]^{2},
\end{equation}
where $d=10$, $\overline{T}$ represents the complex conjugate of the target field, and $\cdot$ denotes the point-wise multiplication. Following the MRAF approach, the cost function is evaluated over the signal region only, as indicated by an asterisk over the summation. Depending on the spatial resolution and/or size of the grid over which phase $\phi$ is defined, the cost function represents a surface in $N^2$-dimensional space for an $N\times N$ grid size. We choose a $512\times512$ grid size to minimize C, which results in the optimization of $512^2$ independent phase values. To achieve this, the gradient of the cost function, $\partial \mathrm{C}/\partial \phi$, is calculated on the multi-dimensional cost surface based on which a conjugate direction is chosen, see Fig.~\ref{flowchart}. While descending along a specific conjugate direction, the cost function is then reduced in finite size steps until a minimum is reached. The resulting phase distribution $\phi_{min}$ is then used to calculate a new gradient and a corresponding conjugate direction for further minimization. This process is iterated until the cost function stagnates. A large constant ($10^d$) in the cost function provides faster convergence while avoiding local minima.

Once the optimization algorithm is finalized, the final phase distribution $\phi_{min}$ is further phase compressed for zero-order suppression~\cite{becker2012ApplOpt}. This is needed due to the limited efficiency of the SLM for higher spatial frequencies of the phase distribution. Such a reduction in efficiency with increasing spatial frequency results in a portion of the field reflected from the SLM without acquiring the calculated phase changes.  The portion of the field that does not experience a phase change presents itself as a zero order in the Fourier plane. The overall phase values can be adjusted to suppress this effect. Finally, after compression, the resulting phase distribution is converted to an 8-bit format image and exported to the SLM.

\section{Spatial auto-correlations and temporal squeezing}\label{Sect:Auto}
\begin{figure}
	\centering
	\includegraphics{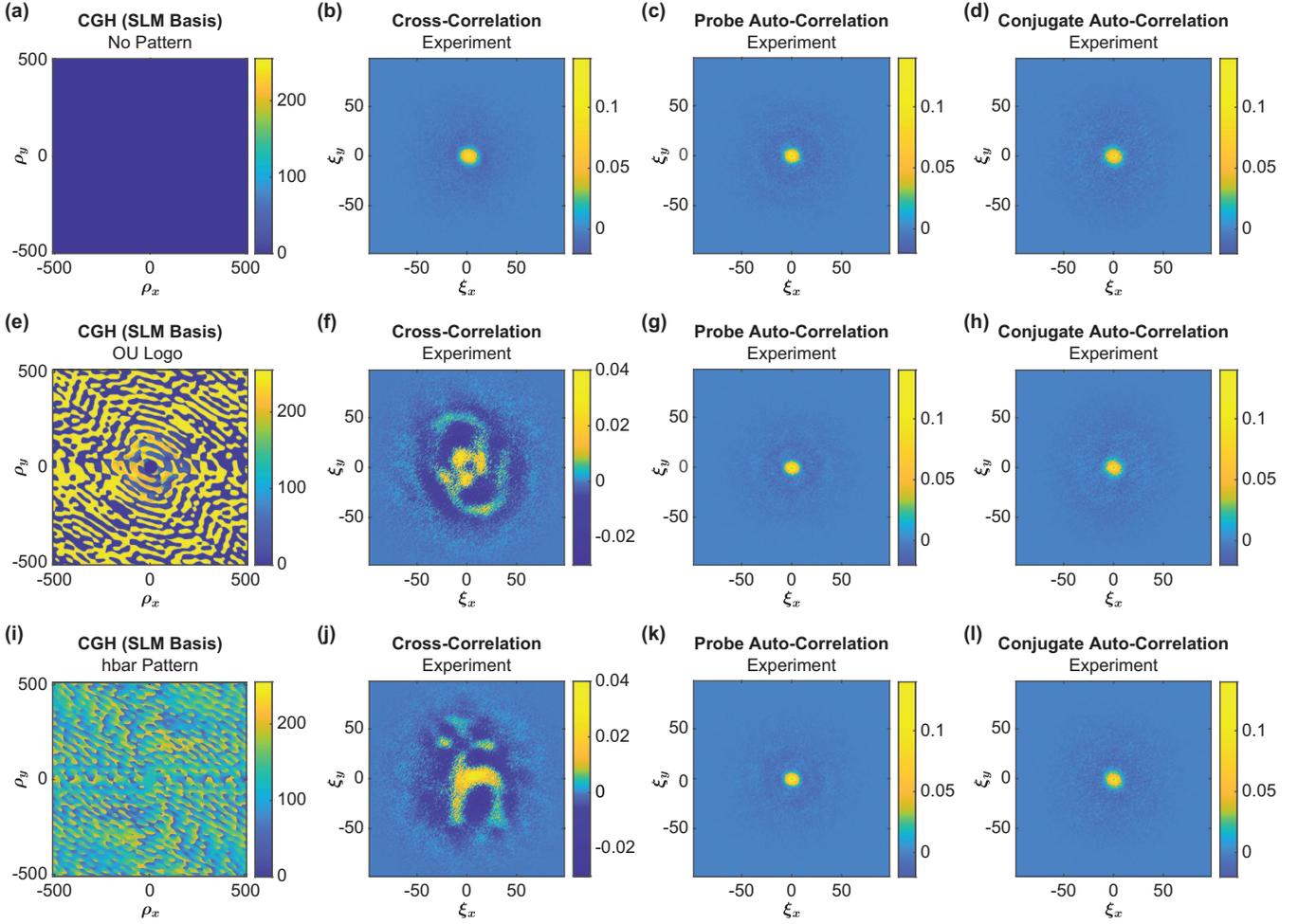}
	\caption{Spatial cross-correlation and auto-correlations for the probe and conjugate for the cases when no pattern is encoded (top row), the OU log is encoded (middle row), and \textcrh ~is encoded (bottom row).  The required phase patterns to encode information on the twin beams for (a) a flat wavefront (no information), (e) the OU logo, and (i) symbol \textcrh ~are transferred to the pump beam through a CGH implemented on an SLM.  The encoded information is extracted through the corresponding cross-correlations (b), (f), and (j), respectively.  On the other hand, the auto-correlations for both the probe and conjugate for all cases remain unchanged even when information is encoded in the distribution of the spatial correlations. Therefore, each beam by itself does not contain the encoded information, which can only be extracted via the spatial cross-correlation between the two beams. All figures, except for the CGH patterns, are shown in the EMCCD pixel basis. The center portion of the auto-correlations was removed as it contains an artificial maximum due to the use of the same images to calculate it.}
	\label{auto_corr}
\end{figure}

In order to be useful for applications in secure communication, it is important for the encoded information to only be accessible through joint measurements of the twin beams and not through individual beam measurements. In order to show that this is the case, we perform the same analysis described in Section~\ref{Sect:data} for each beam individually to evaluate their spatial auto-correlations. The results for the cases in which no pattern is encoded on the pump and for different encoded patterns are shown in Fig.~\ref{auto_corr}.  As can be seen, the auto-correlations for all cases are almost identical and do not reveal the information encoded through the angular spectrum of the pump.

As mentioned in the main text, the fact that the encoded information is not present in the auto-correlations, whose shape is effectively independent of the angular spectrum of the pump (see Fig.~\ref{auto_corr}), is a result of having highly multi-spatial mode twin beams in which a large number of the modes contribute roughly equally to the spatial correlations. In order to show that this is the case, we start with the general wavefunction for spatially multimode  twin beams, which can be written in terms of the spatial eigenmodes of the system as~\cite{Bennink02,Fabre20,Lanning18}
\begin{equation}
\ket{\Psi} = \frac{1}{\sqrt{M}}\sum_{i=1}^{M}\sum_{n=0}^{\infty}\sech(\lambda_{i})\tanh^{n}(\lambda_{i})\ket{\{n\}_{i}}\ket{\{n\}_{i}},
\end{equation}
where $\ket{\{n\}_{i}}$ represents a state with $n$ photons in spatial eigenmode $i$, $\lambda_{i}$ is the degree of squeezing of eigenmode $i$, and $M$ is the number of spatial eigenmodes with $\lambda_{i}\neq0$. The exact spatial profile of the eigenmodes will depend on the angular spectrum of the pump, and  thus will be different for each of the cases considered in Fig.~\ref{auto_corr}.

To calculate the auto-correlation, we first need the reduced density matrix for one of the beams, say the probe.  It is easy to show that the density matrix for the probe beam by itself is given by
\begin{equation}
\rho_{pr}= {\rm Tr}_{c}\{\rho_{pr,c}\}=\frac{1}{M}\sum_{i=1}^{M}\sum_{n=0}^{\infty}\left[\sech(\lambda_{i})\tanh^{n}(\lambda_{i})\right]^{2}\ket{\{n\}_{i}}\bra{\{n\}_{i}}.
\end{equation}
If we now assume that the process generates twin beams with a large number of spatial modes and that they all have roughly the same level of squeezing, as needed for all of them to contribute equally to the spatial correlations, then $\lambda_{i}\equiv\lambda$ such that
\begin{equation}
\rho_{pr}= \frac{1}{M}\sum_{n=0}^{\infty}|A_{n}|^{2}\sum_{i=1}^{M}\ket{\{n\}_{i}}\bra{\{n\}_{i}},
\end{equation}
where we have defined $A_{n}\equiv\sech(\lambda)\tanh^{n}(\lambda)$.

As shown in Section~\ref{Sect:Cross}, for our experimental conditions the fluctuations of the number operator are proportional to the fluctuations of the amplitude quadrature operator.  Thus, in analogy to Eqs.~(\ref{expectation_A}) and~(\ref{expectation_B}), the auto-correlation of the spatial intensity fluctuations of the probe in the far field takes the form
\begin{equation}
\expval{\delta\hat{X}(\bm{x})_{pr}\delta\hat{X}(\bm{x}')_{pr}}=\expval{\hat{X}(\bm{x})_{pr}\hat{X}(\bm{x}')_{pr}},
\end{equation}
where $\hat{X}(\bm{x})=[\hat{a}(\bm{x})+\hat{a}^{\dagger}(\bm{x})]/\sqrt{2}$. To evaluate the auto-correlation, we first expand the field operators in terms of the eigenmodes for the probe, which we denote as $u_{i}(\bm{x})$, to obtain
\begin{equation}
\hat{a}(\bm{x})=\sum_{j=1}^{\infty}u_{j}(\bm{x})\hat{a}_{j},
\end{equation}
where we have taken the spatial dependence out of the field operator $\hat{a}_{j}$, which subtracts a photon from eigenmode $j$. With this expansion we can write
\begin{equation}
\hat{X}(\bm{x})=\frac{1}{\sqrt{2}}\sum_{j=1}^{\infty}\left[u_{j}(\bm{x})\hat{a}_{j}+u_{j}^{\ast}(\bm{x})\hat{a}_{j}^{\dagger}\right],
\end{equation}
which makes it possible to express the spatial auto-correlation after normal ordering as
\begin{align}
\expval{\hat{X}(\bm{x})_{pr}\hat{X}(\bm{x}')_{pr}}&={\rm Tr}\{\rho_{a}\hat{X}(\bm{x})_{pr}\hat{X}(\bm{x}')_{pr}\}\\
&=\frac{1}{2M}\sum_{n=0}^{\infty}|A_{n}|^{2}\sum_{i,j=1}^{\infty}\sum_{k=1}^{M}\left[u_{i}(\bm{x})u_{j}^{\ast}(\bm{x}')\left(\delta_{i,j}+\bra{\{n\}_{k}}\hat{a}_{j}^{\dagger}\hat{a}_{i}\ket{\{n\}_{k}}\right)
+u_{i}^{\ast}(\bm{x})u_{j}(\bm{x}')\bra{\{n\}_{k}}\hat{a}_{i}^{\dagger}\hat{a}_{j}\ket{\{n\}_{k}}\right]\\
&=\frac{1}{2}\sum_{n=0}^{\infty} |A_{n}|^{2}\left[(n+1)\sum_{i=1}^{M}u_{i}(\bm{x})u_{i}^{\ast}(\bm{x}')+n\sum_{i=1}^{M}u_{i}^{\ast}(\bm{x})u_{i}(\bm{x}')\right].
\end{align}
Since the eigenmodes $u_{i}(\bm{x})$ form a complete basis, we have from the closure relation that
\begin{equation}
\sum_{i=1}^{\infty}u_{i}(\bm{x})u_{i}^{\ast}(\bm{x}')=\delta(\bm{x}-\bm{x}').
\end{equation}
Thus, in the limit in which the system generates twin beams with a large number of spatial modes ($M\rightarrow\infty$) such that most of them to contribute equally to the spatial correlations, then
\begin{equation}
\expval{\delta\hat{X}(\bm{x})_{pr}\delta\hat{X}(\bm{x}')_{pr}}\rightarrow\sum_{n=0}^{\infty}\left(n+\frac{1}{2}\right)|A_{n}|^{2}\delta(\bm{x}-\bm{x}').
\end{equation}
That is, the auto-correlations becomes localized. Note that this result is independent of the exact nature of the eigenmodes, which means that in this limit the auto-correlation is independent of the angular spectrum of the pump, as is the case in our experiment (see Fig.~\ref{auto_corr}).  The same procedure can be used to show that the spatial auto-correlation for the conjugate becomes localized in the same limit.

\begin{figure}[h]
	\centering
	\includegraphics{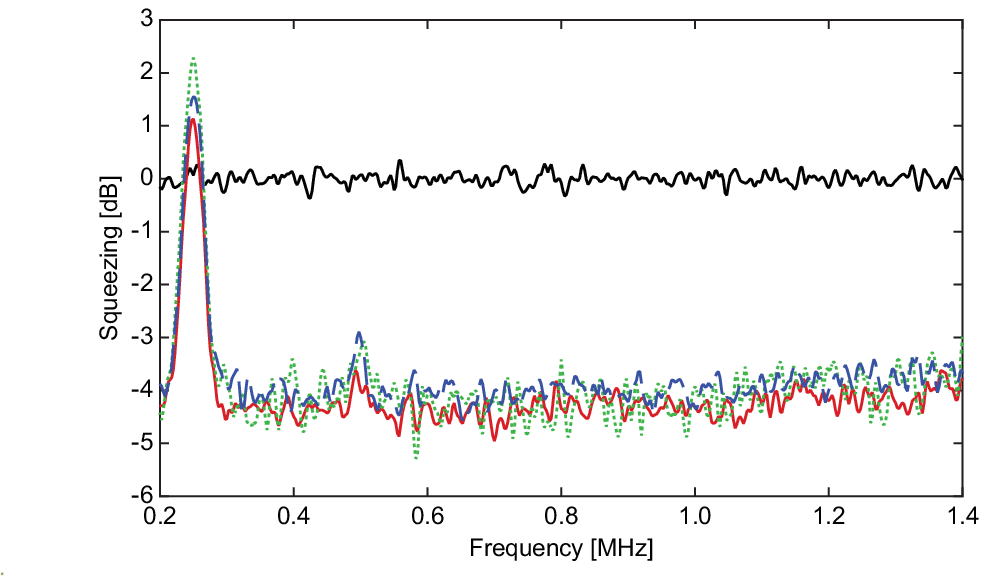}
	\caption{Quantum correlations in the temporal domain are verified through the presence of intensity difference squeezing.  The same level of squeezing is measured, after subtraction of the scattered pump noise, independent of the information encoded in the spatial degree of freedom. The solid black trace shows the shot noise level, while the other traces show  the intensity difference noise when no information is encoded (solid red trace), the OU logo is encoded (dotted green trace), and \textcrh ~is encoded (dashed blue trace). The same levels of temporal squeezing are measured, irrespective of the encoded information, without the need to subtract the scattered pump photons if the additional $^{87}$Rb absorption cell is placed before the photodiodes to absorb the scattered pump, which becomes significant for the case of a structured pump beam.}
	\label{temporal_squeezing}
\end{figure}

Another essential component for the implementation of a secure quantum communication channel using twin beams is for the temporal quantum correlations between the two modes to be preserved even when information is encoded in the distribution of their spatial correlations. To verify that this is the case, we perform temporal intensity difference squeezing measurements by bypassing the EMCCD camera and detecting the bright probe and conjugate fields with photodiodes to perform an intensity difference detection. As can be seen in Fig.~\ref{temporal_squeezing}, the level of temporal intensity difference squeezing is preserved even when the angular spectrum of the pump is modified.  It is important to note that for the results shown in  Fig.~\ref{temporal_squeezing}, we have subtracted the noise from the scattered pump photons, which can become significant for a structured pump.  However, we have verified that if we place the additional isotropically pure $^{87}$Rb cell to filter out the unwanted scattered pump photons in front of the balanced detection system we measure the same level of intensity-difference squeezing without the need of subtracting the background pump noise.  These results show that the degree of temporal quantum correlations is not affected by the information encoded in the spatial degree of freedom.

\end{document}